\documentclass[graybox]{svmult}
\pdfoutput=1

\usepackage{mathptmx}       
\usepackage{helvet}         
\usepackage{courier}        
\usepackage{bm}
\usepackage{amssymb}
\usepackage{type1cm}        
\usepackage{makeidx}         
\usepackage{graphicx}        
\usepackage{multicol}        
\usepackage[bottom]{footmisc}

\def\apj{ApJ}
   
\def\aapr{A\&A Rev.}

\def\apjl{ApJL}

\def\pre{Phys. Rev. E}

\def\na{New Astron.}
\def\aap{A\&A}
\def\araa{ARA\&A}
\def\mnras{MNRAS}

\def\jfm{J. Fluid Mech.}

\newcommand{\code}[1]{\textsf{#1}}
\newcommand{\adp}{\code{ADPDIS3D}}

\makeindex             

\begin{document}

\title*{Energy Transfer and Spectra in Simulations of Two-dimensional Compressible Turbulence}
\author{Alexei G. Kritsuk}
\titlerunning{Energy Spectra in 2D Compressible Turbulence}
\authorrunning{A. G. Kritsuk}
\institute{A. G. Kritsuk \at University of California, San Diego, \email{akritsuk@ucsd.edu}}
%
%
\maketitle
\vspace{-3.6cm}
\begin{figure}[h]
\centering
\includegraphics[scale=.1]{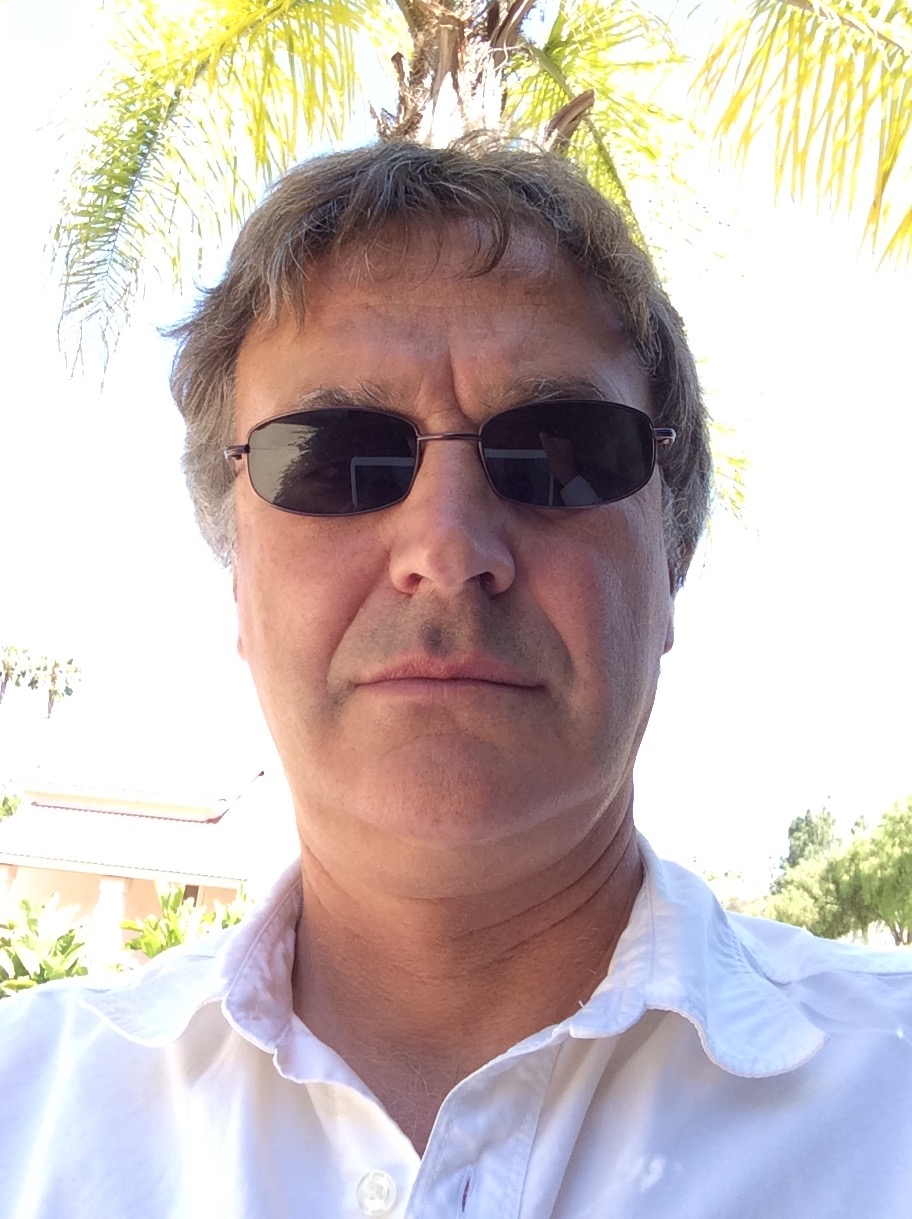}
\end{figure}

\abstract*{
We present results of high-resolution numerical simulations of compressible 2D turbulence forced at intermediate spatial scales with a solenoidal white-in-time external acceleration. A case with an isothermal equation of state, low energy injection rate, and turbulent Mach number $M\approx0.34$ without energy condensate is studied in detail. Analysis of energy spectra and fluxes shows that the classical dual-cascade picture familiar from the incompressible case is substantially modified by compressibility effects. While the small-scale direct enstrophy cascade remains largely intact, a large-scale energy flux loop forms with the direct acoustic energy cascade compensating for the inverse transfer of solenoidal kinetic energy. At small scales, the direct enstrophy and acoustic energy cascades are fully decoupled at small Mach numbers and hence the corresponding spectral energy slopes comply with theoretical predictions, as expected. At large scales, dispersion of acoustic waves on vortices softens the dilatational velocity spectrum, while the pseudo-sound component of the potential energy associated with coherent vortices steepens the potential energy spectrum.
}

\section{Introduction}
Interstellar turbulence \cite{elmegreen.04} is believed to play an important role regulating star formation \cite{mckee.07} in molecular clouds \cite{hennebelle.12}. However, our understanding of large-scale energy cycle in the interstellar medium (ISM) of disk-like galaxies remains incomplete. In particular, it is unclear how the energy injected in the ISM by stellar feedback and gravitational instabilities at scales comparable to the disk scale height $h$ cascades to larger and smaller scales, shaping the structure and global stability of interstellar clouds. The phenomenology of such split energy cascade in quasi-two-dimensional turbulent thin layers of incompressible fluid has been discussed in \cite{boffetta..09,musacchio.17}. While the feasibility of inverse energy transfer in the galactic context 
has attracted some attention \cite{elmegreen.04,bournaud....10}, compressibility effects have never been studied quantitatively in sufficient detail. An important observational signature of the inverse energy cascade that can be verified numerically is the scaling of the column density spectrum, which exhibits a break at $\sim h^{-1}$ in a number of nearby disk-like galaxies observed face-on \cite{dutta...08,dutta...09,block...10,combes12,zhang..12,dutta...13}. Using numerical simulations of two-dimensional (2D) compressible turbulence, we recently demonstrated that the inverse cascade is truncated at turbulent Mach numbers approaching unity, when vortices get destabilized due to acoustic emission \cite{falkovich.17}. The acoustic vortex instability \cite{klyatskin66,broadbent.79,kopev.83,naugolnykh14} ultimately provides for a direct acoustic energy cascade, closing the energy flux loop above the injection scale \cite{falkovich.17}.

In this communication we further detail energy transfer across scales in 2D, using numerical simulations of forced isothermal turbulence in a so-called dual-cascade setting. A high-order accurate low-dissipation numerical method provides enough scale separation to resolve both large- and small-scale 2D cascades on a $16,384^2$ grid. Energy transfer is analyzed in spectral space using our new formalism for compressible turbulence developed in \cite{banerjee.17} and generalized to magnetohydrodynamics in~\cite{banerjee.18}.

\section{Numerics}
\label{sec:2}
We carried out implicit large eddy simulations (ILES) of compressible turbulence in a square periodic domain $L\times L$
covered with a uniform Cartesian grid of $N\times N$ points. The system is governed by the compressible Euler equations 
 {
\begin{eqnarray}
\partial_t \rho+\bm\nabla\cdot(\rho\bm u)& =&0,\label{mas}\\
\partial_t (\rho \bm u)+{\bf \nabla\cdot}\left(\rho\bm u\bm u + p{\bf I}\right)&=&\bm f,\label{mome}\\
\partial_t {\cal E}+{\bf \nabla}\cdot\left[\left({\cal E}+p\right)\bm u\right]& = &{\bm u\cdot \bm f},\label{ener}
\end{eqnarray}
where $\rho$ is the density, $\bm u$ -- velocity, $p$ -- pressure, and ${\cal E}=\rho (u^2/2+e)$ -- total energy density, ${\bf I}=\{\delta_{ij}\}_{i,j=1}^2$ -- identity matrix.}
A solenoidal, white-in-time random external force per unit mass $\bm a=\bm f/\rho$ is applied at an intermediate pumping scale $\lambda_f=2\pi/ k_f$. The system is closed by an ideal gas equation of state  {$p=(\gamma-1)\rho e$} with the ratio of specific heats set very close to unity, $\gamma\equiv c_p/c_v=1.001$, e.g.,~\cite{k07}.
The dimensionless units are chosen so that  the box size $L=1$, the mean density $\rho_0=1$, and the speed of sound $c_{\rm s,0}=1$. The rate of kinetic energy injection by the forcing is relatively small, $\varepsilon_f\in[0.001,0.01]$, as substantially higher rates would inhibit the inverse cascade with most of the added energy dissipated in shocks right at the injection scale. The force correlation time is $\sim10^{4}$ times shorter than the characteristic vortex turn-over time at $\lambda_f$, $\tau_f=\rho_0^{1/3}\lambda_f^{2/3}\varepsilon_f^{-1/3}$. In this formulation, each case is fully defined by three input parameters $(N,\varepsilon_f, \lambda_f)$, see Table~I. 

While we computed about a dozen different cases, here we shall focus only on the two highest resolution weakly forced cases A and B. Case A was evolved through $t_{\rm end}=450$ box-crossing times $\tau=L/c_{\rm s,0}$ with the piecewise parabolic method \cite{colella.84}, reaching the turbulent Mach number $M\approx0.54$. We distinguish the following evolutionary stages: (i) a quasi-incompressible regime with linear total energy growth at a rate $E\equiv\int{\cal E}d\bm x\sim0.92\varepsilon_f t$ at $t\in[0,50]$ and Mach numbers $M\in[0,0.25]$, (ii) a weakly compressible turbulence regime with shocklets at $t\in[50,158]$ and $M\in[0.25,0.3]$, with $E\propto 0.35\varepsilon_f t$, (iii) an energy condensation event at $t\approx158$, marking a fully developed inverse energy cascade, followed by (iv) further growth of the condensate at $t\in[158,380]$ still with $E\propto 0.35\varepsilon_f t$, and (v) energy saturation at $t\in[380,450]$ and $M\sim0.54$ \cite{falkovich.17}.

Case B was restarted from case A at $t=124$ after doubling the inverse cascade range by combining 4 identical boxes into one larger square box covered with a $16,384^2$ grid. We then evolved case B for 26 box-crossing times to get rid of all transients associated with the restart and to further develop the inverse cascade in the new enlarged domain. Finally, we evolved the case for $\Delta t=30\tau$ at $M\sim0.33$ and collected 600 data snapshots at $t\in[150,180]$ to study turbulence statistics. For this simulation, we used a more accurate method described below, which allowed us to double the spectral bandwidth in the inertial range of the direct enstrophy cascade (i.e. below the energy injection scale). Hence, we essentially doubled the extent of inertial ranges for both incompressible cascades compared to case A.
\begin{table}[t]
\caption{Simulations and parameters}
\label{tab:1}  
\begin{tabular}{p{1.2cm}p{1.5cm}p{1.3cm}p{1.2cm}p{1.2cm}p{1.2cm}p{1.2cm}p{1.2cm}p{0.6cm}}
\hline\noalign{\smallskip}
Case & $N$ & $\lambda_f$ & $\varepsilon_f$ & $t_{\rm start}$ & $t_{\rm cnd}$ & $t_{\rm sat}$ & $t_{\rm end}$ & $M_{\rm end}$ \\
\noalign{\smallskip}\svhline\noalign{\smallskip}
A & 8192 & 0.012 & 0.001& 0 & 158 & 390 & 450 & 0.52\\
B & 16\,384 & 0.006 & 0.001& 124 & --- & --- &180 & 0.34\\
\noalign{\smallskip}\hline\noalign{\smallskip}
\end{tabular}
\end{table}

To evolve case B, we used a variable high-order 3D solver developed for problems involving turbulence with 
strong shocks and density variations at flow speeds that range from nearly incompressible to hypersonic \cite{kotov16,wang....11} and implemented in the \adp\ code.
Our production runs used an optimal subset of numerical methods, which includes:  (i) 8th-order-accurate central spatial base scheme that employs a split form of the inviscid flux derivative for better numerical stability \cite{ducros.....00}; 
(ii) non-linear Ducros et al. sensor \cite{ducros......99} to filter the solution and provide extra dissipation where needed, using a dissipative portion of the 7th-order WENO scheme and limiting the use of numerical dissipation away from discontinuities with a control parameter, distinguishing shocks from vortical flow types; and (iii) 4th-order Runge-Kutta time integration. 

\section{Results}
\label{sec:3}

\begin{figure}[t]
\sidecaption[t]
\includegraphics[scale=.59]{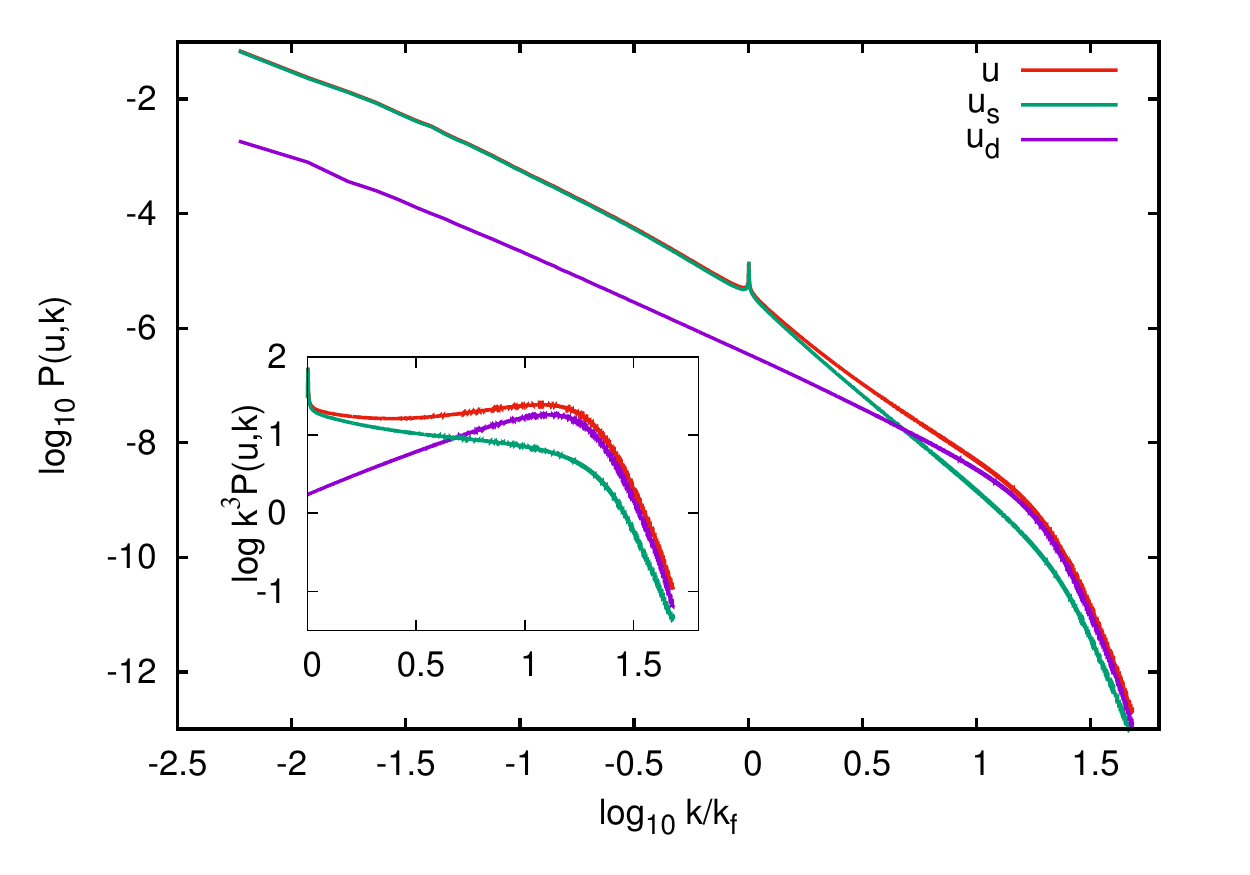}
\caption{Power spectra of the velocity $\bm u$ (red) and its solenoidal $\bm u_s$ (green) and dilatational $\bm u_d$ (purple) components. The solenoidal component dominates at large scales, $k\lesssim k_f$, while the dilatational one dominates at small scales, $k\gg k_f$, resulting in a `bottleneck' in the velocity spectrum, see the inset for $k^3$-compensated small-scale spectra.}
\label{fig:helm}  
\end{figure}

Time-averaged velocity power spectra for case B are shown in Fig.~\ref{fig:helm}.
The spectrum is defined by $P(\bm u,k)\equiv\int|\widehat{\bm u}(\bm \kappa)|^2\delta(k-|\bm\kappa|)d\bm\kappa$, where $\;\widehat{\bm u}(\bm k)\;$ denotes the Fourier transform of the velocity $\bm u(\bm x)$ and $\delta(k)$ is the Dirac delta function. Averaging over a short period of time $\Delta t=30\tau$ at $t>150$ is justified because at $\varepsilon_f=0.001$ the energy growth $E(t)\propto 0.33\varepsilon_f t$ is slow and only affects the tip of the spectrum at the lowest wave numbers, while the rest of the spectrum remains statistically stationary. As can be seen in Fig.~\ref{fig:helm}, the spectral slopes deviate  from classical predictions for incompressible turbulence in two dimensions \cite{kraichnan67,kraichnan71}. 

With a well-resolved stochastic forcing, we obtain $P(\bm u,k)\propto k^{-2}$ at $k<k_f$ instead of $P(\bm u,k)\propto k^{-5/3}$. The same scaling was measured at similar Mach numbers in case A, just before the energy condensation occurred \cite{falkovich.17}. 
At $k>k_f$, where one would normally expect to see $P(\bm u,k)\propto k^{-3}$, the spectrum does not show any clear power-law scaling range, even though the numerical method used in case B is sufficiently accurate to resolve an inertial range.

To discuss the origin of these deviations in compressible turbulence, we use Helmholtz decomposition $\bm u=\bm u_s+\bm u_d$, separating solenoidal  $\bm u_s$ and dilatational $\bm u_d$ velocity components. The decomposed spectra $P(\bm u_s,k)$ and $P(\bm u_d,k)$ are also shown in Fig.~\ref{fig:helm}. The solenoidal component $\bm u_s$ clearly dominates over the dilatational one at all wave numbers, except for $k\gtrsim5k_f$. A local peak in $P(\bm u_s,k)$ at $k_f$ is associated with the forcing, while $P(\bm u_d,k)$ does not show any feature at $k_f$ because the external acceleration $\bm a_f=\bm f/\rho$ is divergence-free. 

\begin{figure}[b]
\sidecaption[b]
\includegraphics[scale=.59]{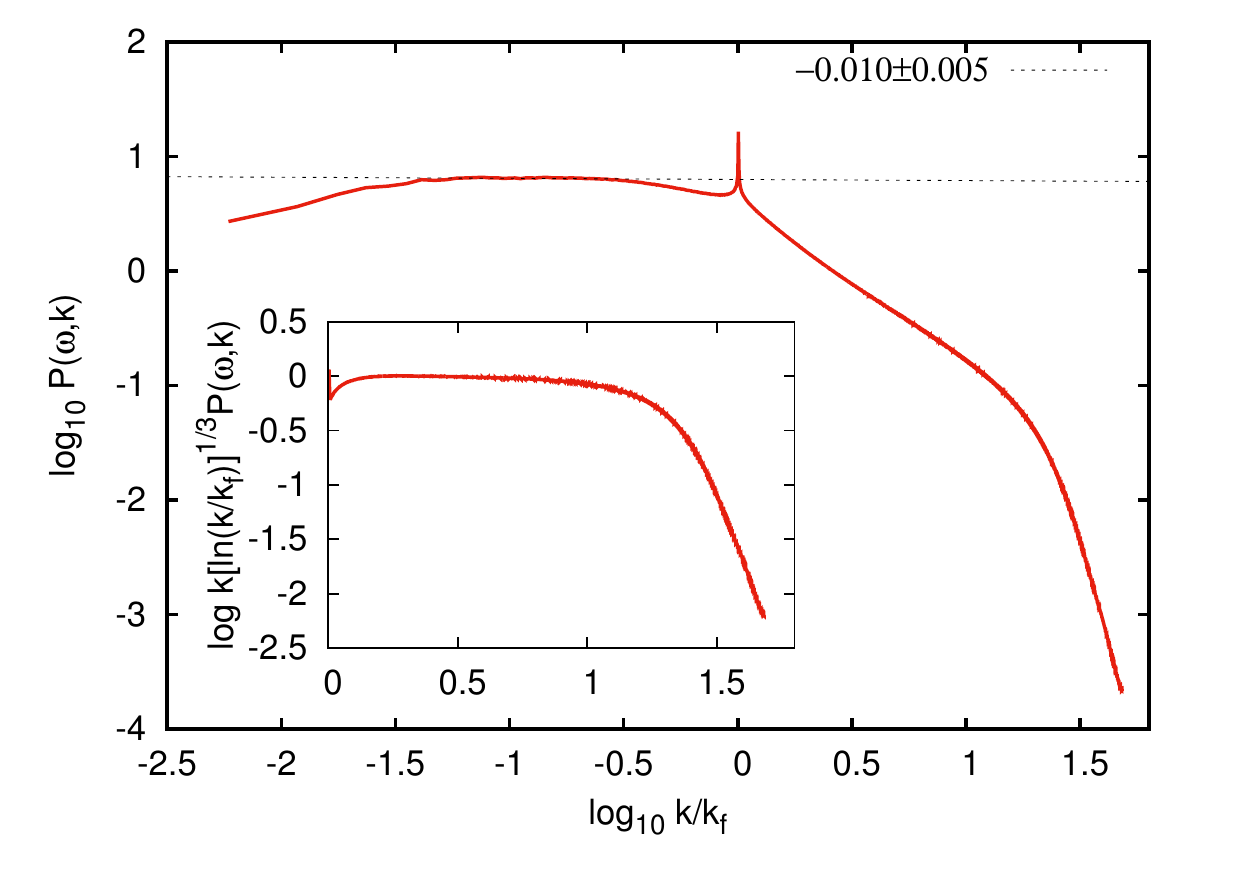}
\caption{Vorticity power spectrum is essentially flat at $k<k_f$, $P(\omega,k)\propto k^0$, consistent with $P(\bm u_s,k)\propto k^{-2}$. At $k>k_f$, the spectrum closely follows Kraichnan's prediction $P(\omega,k)\propto k^{-1}[\ln{(k/k_f)}]^{-1/3}$ \cite{kraichnan71}, see compensated spectrum in the inset.}
\label{fig:vort}  
\end{figure}

\begin{figure}[t]
\sidecaption[t]
\includegraphics[scale=.59]{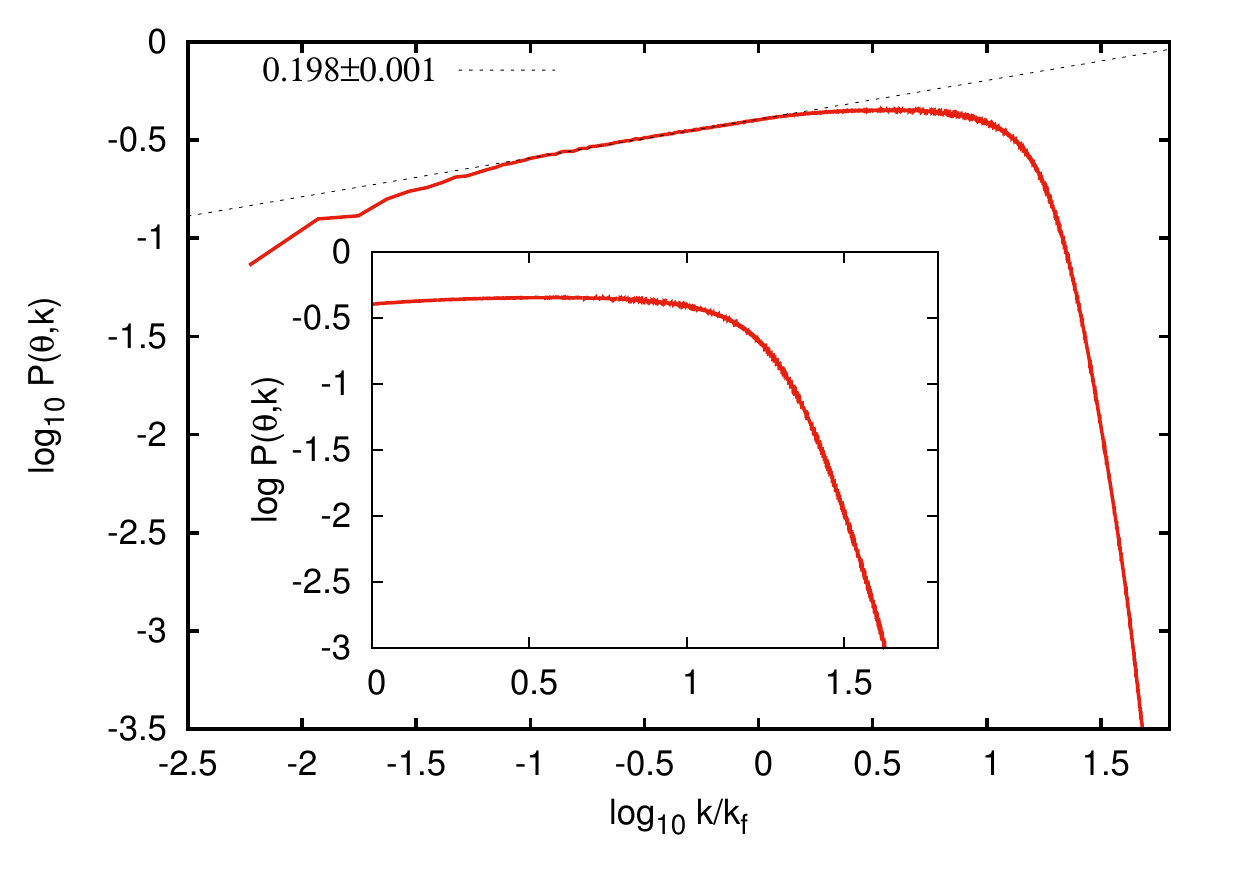}
\caption{Power spectrum of the velocity divergence $\theta\equiv\bm\nabla\cdot\bm u$. At large scales $P(\theta,k)\propto k^{1/5}$, while at $k>k_f$ the spectrum scales approximately as $P(\theta,k)\propto k^{0}$, reflecting $P(\bm u_d,k)=k^{-2}P(\theta,k)\propto k^{-2}$, see the inset.}
\label{fig:div} 
\end{figure}

To detail the velocity scaling further, we show spectra of the vorticity $\bm\omega\equiv\bm\nabla\times\bm u=\bm\nabla\times\bm u_s$ and dilatation $\theta\equiv\bm\nabla\cdot\bm u=\bm\nabla\cdot\bm u_d$ in Figs.~\ref{fig:vort} and \ref{fig:div}, respectively. 
Above the injection scale, at $k/k_f\in[0.03,0.3]$, the vorticity spectrum is flat $P(\bm\omega,k)\propto k^0$, corresponding to $P(\bm u_s,k)\propto k^{-2}$. There is a slight deep in the spectrum just above the injection scale at $k/k_f\in[0.3,1]$, where a small fraction of pumped up solenoidal kinetic energy is being converted into acoustic energy. 

At $k>k_f$, the vorticity spectrum is steeper than $k^{-1}$ and hence $P(\bm u_s,k)$ is steeper than $k^{-3}$, as can also be seen in the inset of Fig.~\ref{fig:helm}. The logarithmic correction $[\ln(k/k_f)]^{-1/3}$, however, is sufficient to have a compensated spectrum approximately flat for about a decade in $k$ (see inset in Fig.~\ref{fig:vort}).
The solenoidal velocity spectrum, thus, closely follows Kraichnan's prediction \cite{kraichnan71} for the direct enstrophy cascade in incompressible 2D turbulence, i.e. $P(\bm u_s,k)\propto k^{-3}[\ln(k/k_f)]^{-1/3}$ at $k>k_f$. It is worth noting that smooth flows in ideal 2D compressible hydrodynamics conserve the potential vorticity $\bm\omega/\rho$ of any streamline, but when compressibility is small, the enstrophy cascade persists, much as in the
incompressible case \cite{falkovich.17}.

\begin{figure}[t]
\sidecaption[t]
\includegraphics[scale=.59]{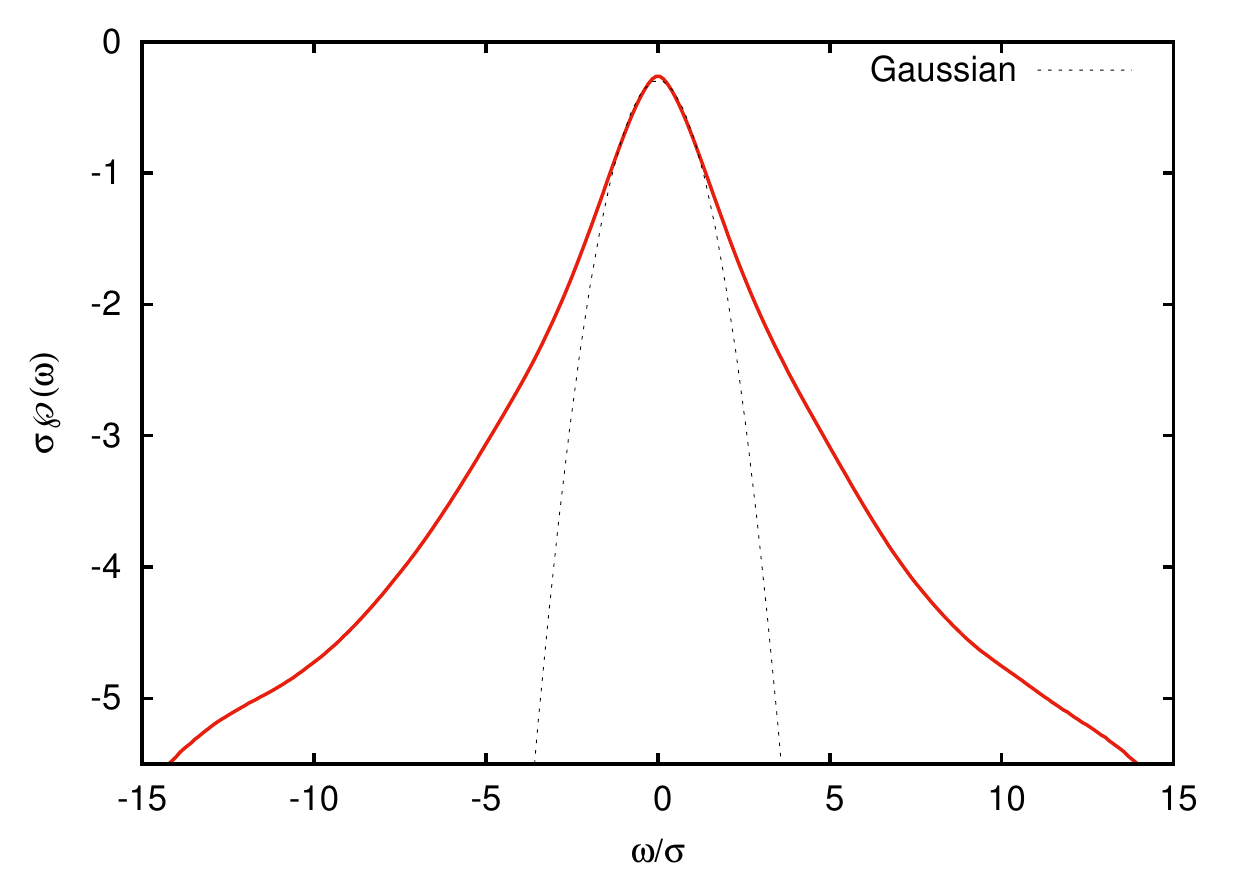}
\caption{The vorticity PDF is strongly non-Gaussian, reflecting the presence of coherent vortices. Vorticity $\omega$ is normalized by its rms fluctuations $\sigma$.}
\label{fig:vorpdf} 
\end{figure}

At the same time, the spectrum of solenoidal velocity $P(\bm u_s,k)\propto k^{-2}$ is substantially steeper than $k^{-5/3}$ at $k<k_f$. Similar spectral slopes were previously seen in 2D simulations of incompressible turbulence with stochastic forcing in which the forcing scale $\lambda_f$ is sufficiently well resolved and large-scale friction is not included \cite{scott07}. The steep spectra were associated with the emergence of coherent vortices, which usually coexist with the inverse energy cascade in 2D \cite{burgess..15}. The vortices do not form if $\lambda_f$ is unresolved; they also may get destabilized if a stationary external force is used \cite{mizuta..13} or if the order of hypodissipation is small enough \cite{boffetta..00}. The presence of coherent vortices is reflected in the strongly non-Gaussian shape of single-point vorticity PDF (see Fig.~\ref{fig:vorpdf} and Ref.~\cite{scott07}). Moreover, scaling exponents $\zeta_p$ of the transverse solenoidal velocity structure functions $|\Delta \bm u_s^{\perp}(\ell)|^p\propto\ell^{\zeta_p}$ also show anomalies with $\zeta_p>p/3$ at $p<3$ and saturate at $\zeta_p\approx1$ for order $p\in[3,6]$ due to the presence of vortices \cite{wang.17}, cf.~\cite{boffetta..00}. Overall, the emerging population of vortices appears to substantially control the dynamics of 2D turbulence in our cases A and B. 

Let us now consider the spectrum of dilatation (Fig.~\ref{fig:div}), which has a small positive slope $P(\theta,k)\propto k^{0.2}$ at $k<k_f$ and then flattens to $P(\theta,k)\propto k^0$ at $k>k_f$. The dilatational velocity spectrum at small scales $P(\bm u_d,k)\propto k^{-2}$ is consistent with theoretical prediction for the potential velocity component in acoustic turbulence by Kadomtsev and Petviashvili \cite{kadomtsev.73}. Indeed, at small Mach numbers, the direct acoustic energy and enstrophy cascades proceed independently of each other at $k\gtrsim k_f$. 

At $k<k_f$, we observe a slightly more shallow spectrum of dilatational velocity $P(\bm u_d,k)\propto k^{-1.8}$. Due to dispersion of acoustic waves on large-scale coherent vortices, the slope is expected to lie approximately half-way between $-2$ (the case of purely potential velocity \cite{kadomtsev.73}) and $-11/7\approx-1.57$ (as suggested for 2D in Ref.~\cite{elsasser.76} based on arguments similar to those advanced for 3D acoustic turbulence by Zakharov and Sagdeev \cite{zakharov.70}, see also \cite{moiseev...77,lvov.78a,lvov.78b,lvov.81}).

\begin{figure}[b]
\sidecaption[b]
\includegraphics[scale=.59]{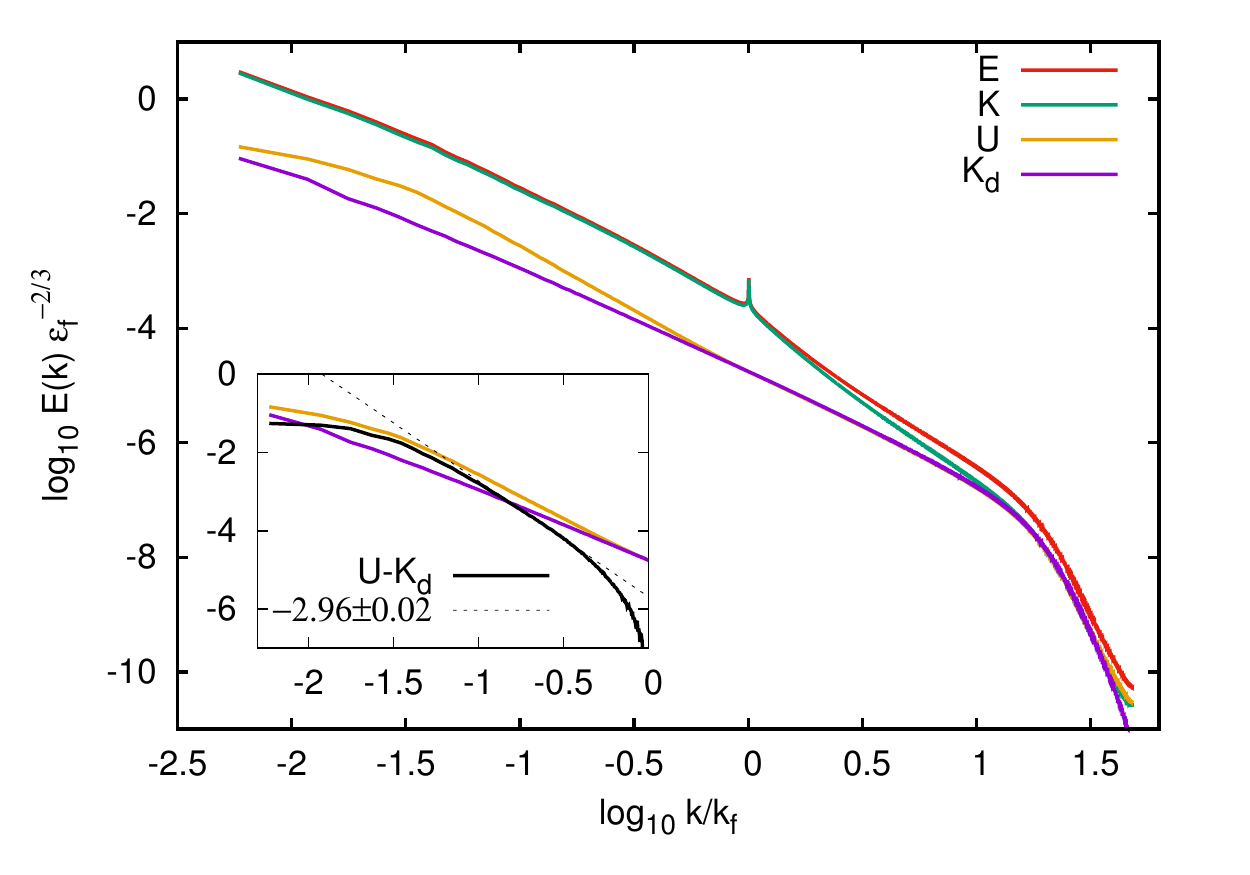}
\caption{Total $E(k)$ (red), kinetic $K(k)$ (green), and potential $U(k)$ (brown) spectral energy densities. Also shown is the dilatational component $K_d(k)$ (purple) of the kinetic energy in detailed equipartition with the potential one $K_d(k)\approx U(k)$ at $k\gtrsim k_f$.}
\label{fig:copow}  
\end{figure}

Besides the velocity spectra, it is worth inspecting the spectral densities of kinetic and potential energy, $K(k)=P(\rho\bm u, \bm u; k)/2$ and $U(k)=P(\rho, e; k)/2+U_0\delta(k)/2$, respectively (here $U_0\equiv\int\rho ed\bm x$). Indeed, the total energy $E=\int_0^{\infty}E(k)dk=K+U=\int_0^{\infty}[K(k)+U(k)]dk$ is an ideal invariant of the isothermal system. Following Ref.~\cite{banerjee.17}, we define the spectral densities as cospectra $P(\bm a,\bm b;k)\equiv\int[\widehat{\bm a}(\bm \kappa)\cdot\widehat{\bm b}^*(\bm \kappa)+\widehat{\bm a}^*(\bm \kappa)\cdot\widehat{\bm b}(\bm \kappa)]\delta(k-|\bm\kappa|)d\bm\kappa/2$, with $\bm a=\rho\bm u$, $\bm b=\bm u$ in case of the kinetic energy and $a=\rho$, $b=e$ for the potential energy. While these generic definitions are valid for arbitrary degree of compressibility, in the Mach number regimes realized in case B, the kinetic energy spectra can be reasonably well approximated by $K(k)\approx\rho_0P(\bm u,k)/2$ at all resolved wave numbers. Likewise, the kinetic energy spectral density can be approximately decomposed into solenoidal and dilatational parts $K(k)\approx K_s(k)+K_d(k)$, where $K_s(k)=\rho_0P(\bm u_s,k)/2$ and $K_d(k)=\rho_0P(\bm u_d,k)/2$.

Figure~\ref{fig:copow} shows the relevant spectral energy densities: total $E(k)$, kinetic $K(k)$, potential $U(k)$, and dilatational kinetic $K_d(k)$. Overall, these look similar to the corresponding velocity spectra, except for $U(k)$, which is new. One can clearly see the detailed acoustic energy equipartition $U(k)\approx K_d(k)$ at $k\gtrsim k_f$ \cite{sarkar...91,banerjee.17}. However, at large scales, the presence of coherent vortices breaks this equipartition, as pseudo-sound component of $U(k)$ associated with the vortices makes the potential energy exceed $K_d(k)$ at $k<k_f$. The inset in Fig.~\ref{fig:copow} details the pseudo-sound contribution $U(k)-K_d(k)$ shown in black, which scales approximately as $k^{-3}$. This scaling can be readily derived, assuming that centrifugal force is balanced by the pressure gradient in coherent vortices and $P(\bm u,k)\propto k^{-2}$. We thus see a large-scale excess of $U(k)$ as another (purely compressible) signature of the presence of coherent vortices in the inertial range of inverse energy cascade.

\begin{figure}[t]
\sidecaption[t]
\includegraphics[scale=.59]{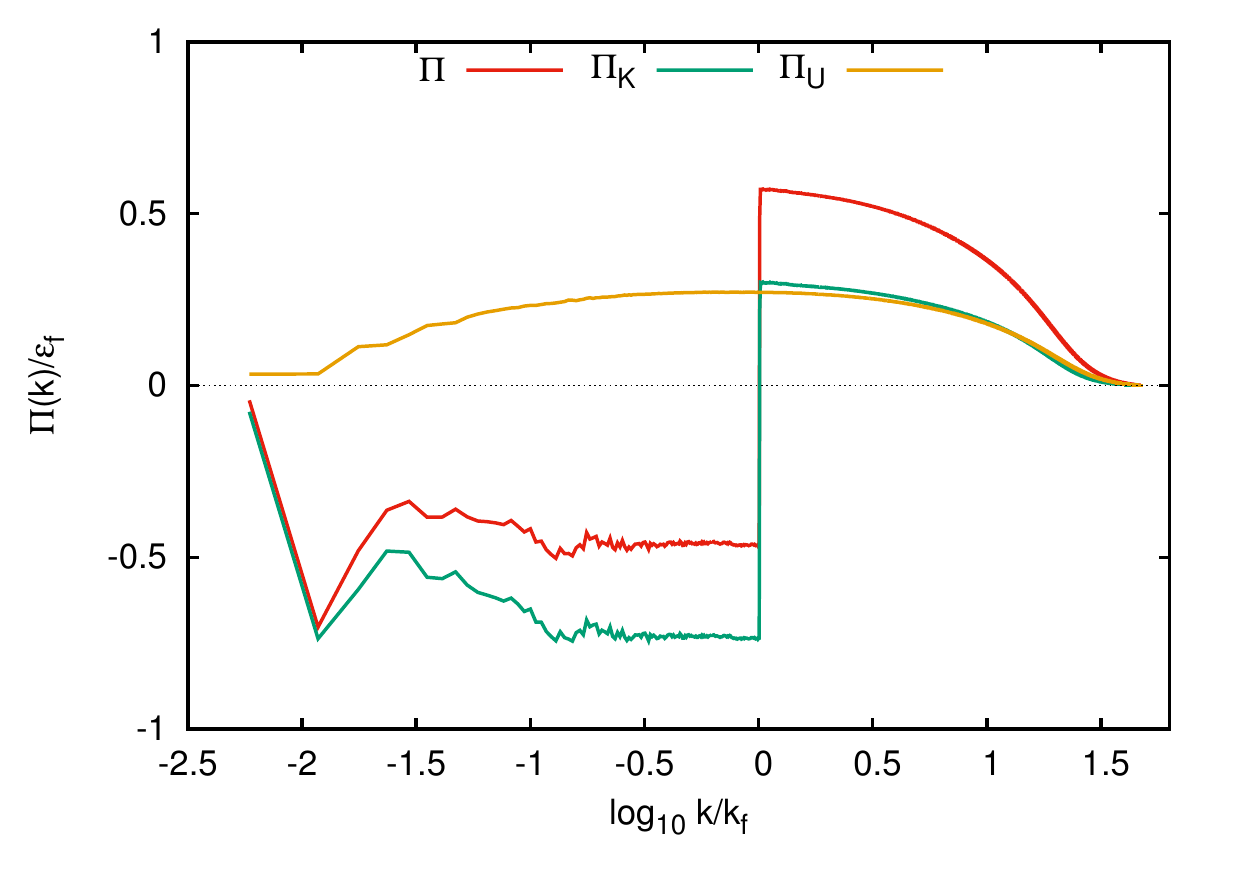}
\caption{Net $\Pi(k)$ (red), kinetic $\Pi_K(k)$ (green), and potential $\Pi_U(k)$ (brown) spectral energy fluxes normalized by the kinetic energy injection rate $\varepsilon_f$. Note that the total energy $E(t)$ is not statistically stationary and continues to grow at a rate of approximately $0.33\varepsilon_f$ within the time-averaging interval $t\in[150,180]$.}
\label{fig:flux}   
\end{figure}

Finally, spectral energy fluxes computed using the formalism developed in Ref.~\cite{banerjee.17} are shown in Fig.~\ref{fig:flux}. The kinetic $\Pi_K(k)$ (green) and potential $\Pi_U(k)$ (brown) energy fluxes form a flux loop at $k<k_f$, as solenoidal kinetic energy inversely cascades to large scales, where it gets converted into acoustic energy, and then directly cascades to small scales \cite{falkovich.17}. The net total energy flux $\Pi(k)=\Pi_K(k)+\Pi_U(k)$ (red) is split in two roughly equal parts: one cascading inversely to feed the continuing energy growth of the system, and another cascading directly to rid the system of the excessive acoustic noise. In the mean time, shock dissipation actively drains the kinetic energy of the isothermal system across scales. It is worth noting that the kinetic and potential components of the net flux are comparable, while the kinetic-to-potential energy ratio generally oscillates around 10\%. Thus, even small compressibility can alter or even reverse the energy transfer across scales. Another remark due here is on the notion of `kinetic energy cascade' sometimes used in compressible turbulence, even though $K$ is not an invariant of the dynamics, e.g., \cite{aluie..12}. Our 2D case provides a curious counterexample, as the solenoidal and dilatational components of the kinetic energy cascade in opposite directions at $k<k_f$.

\section{Summary}
\label{sec:4}
We presented results of high-resolution numerical simulations of compressible 2D turbulence forced at intermediate spatial scales with a solenoidal white-in-time external acceleration. We studied in detail a case with an isothermal equation of state, low energy injection rate, and turbulent Mach number $M\approx0.34$ without energy condensate. Our analysis of energy spectra and fluxes shows that the classical dual-cascade picture familiar from the incompressible case is substantially modified by compressibility effects. While the small-scale direct enstrophy cascade remains largely intact, a large-scale energy flux loop forms with the direct acoustic energy cascade compensating for the inverse transfer of solenoidal kinetic energy. At small scales, the direct enstrophy and acoustic energy cascades are fully decoupled at low Mach numbers, and hence the corresponding spectral energy slopes comply with theoretical predictions \cite{kraichnan71,kadomtsev.73}, as expected. At large scales, dispersion of acoustic waves on vortices softens the dilatational velocity spectrum \cite{zakharov.70,moiseev...77}, while pseudo-sound component of the potential energy associated with coherent vortices steepens the potential energy spectrum.

\begin{acknowledgement}
This research was supported in part by the National Science Foundation through Grant
No.~AST-1412271  
as well as through XSEDE allocation MCA07S014 on {\em Stampede-1/2} at TACC (production runs) and on {\em Comet} at SDSC (data analysis).
\end{acknowledgement}


\end{document}